\documentclass[12pt,onecolumn,conference]{IEEEtran}

\usepackage{latexsym}
\usepackage{amssymb}
\usepackage{bm}
\usepackage[dvips]{graphics}
\usepackage{graphicx}
\usepackage{psfrag}
\usepackage{amsfonts,amsmath,amssymb,hyperref}

\usepackage{dsfont,color,subfigure}

\DeclareMathOperator*{\argmin}{arg\,min}

\usepackage{xspace}
\usepackage{bbm}

\newcommand{\Ec}{\mathcal{E}}

\newcommand{\Sc}{\mathcal{S}}

\newcommand{\Xc}{\mathcal{X}}

\def\a{\alpha}
\def\b{\beta}

\def\l{\lambda}

\DeclareMathOperator\E{E}
\let\P\relax
\DeclareMathOperator\P{P}

\def\textiid{i.i.d.\@\xspace}
\newcommand\iid{\ifmmode\text{ i.i.d. } \else \textiid \fi}

\newtheorem{theorem}{Theorem}
\newtheorem{lemma}{Lemma}

\begin{document}

\title{An Implementable Scheme for Universal Lossy  Compression of Discrete Markov Sources}

\author{\authorblockN{\small{Shirin Jalali},\authorrefmark{1} Andrea Montanari\authorrefmark{1} and Tsachy Weissman\authorrefmark{1}\authorrefmark{2},}
\authorblockA{\authorrefmark{1}\small{Department of Electrical
Engineering, Stanford University, Stanford, CA 94305},\\
\authorblockA{\authorrefmark{2}
\small{Department of Electrical Engineering, Technion, Haifa
32000, Israel}\\
\small{\{shjalali, montanar, tsachy\}@stanford.edu}}}}
\maketitle

\newcommand{\p}{\mathds{P}}
\newcommand{\mb}{\mathbf{m}}
\newcommand{\bb}{\mathbf{b}}

\begin{abstract}
We present a new lossy compressor for discrete sources. For
coding a source sequence $x^n$, the encoder starts by assigning
a certain cost to each reconstruction sequence. It then finds
the reconstruction that minimizes this cost and describes it
losslessly to the decoder via a universal lossless compressor.
The cost of a sequence is given by a linear combination of its
empirical probabilities of some order $k+1$ and its distortion
relative to the source sequence. The linear structure of the
cost in the empirical count  matrix allows the encoder to
employ a Viterbi-like algorithm for obtaining the minimizing
reconstruction sequence simply. We identify a choice of
coefficients for the linear combination in the cost function
which ensures that the algorithm universally achieves the
optimum rate-distortion performance of any Markov source in the
limit of large $n$, provided $k$ is increased as $o(\log n)$.
\end{abstract}

\section{Introduction}

Let $\mathbf{X}=\{X_i: i \geq 1\}$ represent a discrete-valued
stationary ergodic process with unknown statistics, and
consider the problem of compressing $\mathbf{X}$ at rate $R$
such that the incurred distortion is minimized. Let
$\mathcal{X}$ and $\hat{\mathcal{X}}$ denote finite source and
reconstruction alphabets respectively. The performance of the
described coding scheme is measured by its average expected
distortion between source and reconstruction blocks, i.e.~
\begin{equation}
D=\E
d_n(X^n,\hat{X}^n)\triangleq\frac{1}{n}\sum\limits_{i=1}^n \E d(X_i,\hat{X}_i),
\end{equation}
where $d:\mathcal{X}\times\mathcal{X}\rightarrow
{\mathds{R}}^+$ is a single-letter distortion measure.  For any
$R\geq0$, the minimum achievable distortion (cf.~\cite{cover}
for exact definition of achievability) is characterized as
\cite{Shannon}, \cite{Gallager}, \cite{book: Berger}
\begin{equation} \label{eq: rate-distortion function}
D(\mathbf{X},R)=\lim\limits_{n\rightarrow\infty}\min\limits_{p(\hat{X}^n|X^n):I(X^n;\hat{X}^n)\leq
R}\E d_n(X^n,\hat{X}^n).
\end{equation}
A sequence of codes at rate $R$ is called universal if for
every stationary ergodic source $\mathbf{X}$ its asymptotic
performance converges to $D(\mathbf{X},R)$, i.e.,
\begin{align}
\limsup\limits_{n\to\infty} \E d_n(X^n,\hat{X}^n)\leq
D(\mathbf{X},R).
\end{align}

For lossless compression where the source is to be recovered
without any errors, there already exist well-known
implementable universal schemes such as Lempel-Ziv coding
\cite{LZ} or arithmetic coding \cite{arith_coding}. In contrast
to the situation of lossless compression,  for $D>0$, there are
no well-known practical schemes that universally achieve the
rate-distortion curve. In recent years, there has been progress
towards designing universal lossy compressor especially in
trying to tune some of the existing universal lossless coders
to work in the lossy case as well \cite{kontoyiannis_1},
\cite{universal_fixed_slope_coder}, \cite{yang_kieffer}. All of
these algorithms are either provably suboptimal, or optimal but
with exponential complexity.

Another approach for lossy compression, which is very
well-studied in the literature and even implemented in JPEG
2000 image compression standard, is Trellis coded quantization,
i.e.~Trellis structured code plus Viterbi encoding
(c.f.~\cite{tutorial lossy compression}, \cite{Gray book} and
references therein). This method is in general suboptimal for
coding sources that have memory \cite{Gray book}. In
\cite{variable rate Trellis}, an algorithm for fixed-slope
Trellis source coding is proposed, and is shown to be able to
get arbitrary close to the rate-distortion curve for
continuous-valued stationary ergodic sources. The proposed
method is efficient in low rate region.

In a recent work \cite{R_D MCMC}, a new implementable algorithm
for lossy compression of discrete-valued stationary  ergodic
sources was proposed. Instead of fixing rate (or distortion)
and minimizing distortion (or rate), the new algorithm fixes
Lagrangian coefficient $\alpha$, and minimizes $R+\a D$. This
is done by assigning energy $\Ec(y^n)$ representing $R+\a D$ to
each possible reconstruction sequence and finding the sequence
that minimizes the cost by simulated annealing. The algorithm
starts by letting $y^n=x^n$, and at each iteration chooses an
index $i\in\{1,\ldots,n\}$ uniformly at random, and
probabilistically changes $y_i$ to some $y\in\hat{\Xc}$ such
that there is a positive probability (which goes to zero as the
number of iterations increases) that the resulting sequence has
higher energy than the original sequence. Allowing the energy
to increase especially at initial steps prevents the algorithm
from being entrapped in a local minimum. It was shown that
using a universal lossless compressor to describe the
reconstruction sequence resulting from this process to the
decoder results in a scheme which is universal in the limit of
many iterations and large block length. The drawback of the
proposed scheme is that although its computational complexity
per iteration is independent of the block length $n$ and linear
in a parameter $k_n=o(\log n)$, there is no useful bound on the
number of iterations required for convergence. In this paper,
inspired by the previous method, we propose yet another
approach for lossy compression of discrete Markov sources which
universally achieves optimum rate-distortion performance for
any discrete Markov source. We start by assigning the same cost
that was defined for each possible reconstruction sequence in
\cite{R_D MCMC}. The cost of each sequence is a linear
combination of two terms: its empirical conditional entropy and
its distance to the source sequence to be coded. We show that
there exists proper linear approximation of the first term such
that minimizing the linearized cost results in the same
performance as minimizing the original cost. But the advantage
is that minimizing the modified cost can be done via Viterbi
algorithm in lieu of simulated annealing which was used for
minimizing the original cost.

The organization of the paper is as follows. In Section
\ref{sec:notation}, we set up the notation, and define the
count matrix and empirical conditional entropy of a sequence.
Section \ref{sec: linearized cost} describes a new coding
scheme for fixed-slope lossy compression which universally
achieves the rate-distortion curve for any discrete Markov
source and \ref{sec: how to choose} describes how to compute
the coefficients required by the algorithm outlined in the
previous section. Section \ref{sec: Viterbi coding} explains
how Viterbi algorithm can be used for implementing the coding
scheme described in Section \ref{sec: linearized cost}. Section
\ref{sec: simulation results} presents some simulations
results, and finally, Section \ref{sec: conclusion} concludes
the paper with a discussion of some future directions.

Proofs that are not presented in the paper will appear in the
full version.

\section{Notations and required definitions}\label{sec:notation}

Let $\Xc$ and $\hat{\Xc}$ denote the source and reconstruction
alphabets respectively. Let matrix
$\mathbf{m}(y^n)\in\mathds{R}^{|\hat{\Xc}|}\times\mathds{R}^{|\hat{\Xc}|^{k}}$
represent $(k+1)^{\rm th}$ order empirical count of $y^n$
defined as
\begin{equation}\label{eq: empirical count matrix}
    m_{\b,\bb}(y^n) = \frac{1}{n} \left| \left\{ 1 \leq i \leq n : y_{i-k}^{i-1} = \bb, y_i=\b]    \right\}
    \right|.
\end{equation}
In \eqref{eq: empirical count matrix}, and throughout we assume
a cyclic convention whereby $y_i \triangleq y_{n+i}$ for $i
\leq 0$. Let $H_k (y^n)$ denote the conditional empirical
entropy of order $k$ induced by $y^n$, i.e.~
\begin{equation}\label{eq: emp cond distribution}
    H_k (y^n) = H(Y_{k+1} | Y^{k}) ,
\end{equation}
where $Y^{k+1}$ on the right hand side of (\ref{eq: emp cond
distribution}) is distributed according to
\begin{equation}\label{eq: empirical distribution}
    \P (Y^{k+1} = [\bb,\b]) = m_{\b,\bb}(y^n),
\end{equation}
where $\b\in\hat{\Xc}$, and $\bb\in\hat{\Xc}^{k}$, and
$[\bb,\b]$ represents the vector made by concatenation of $\bb$
and $\b$. We will use the same notation throughout the paper,
namely, $\b,\b',\ldots\in\hat{\Xc}$, and   $\bb, \bb',
\ldots\in\hat{\Xc}^k$. The conditional empirical entropy in
\eqref{eq: emp cond distribution} can be expressed as a
function of $\mb(y^n)$ as follows
\begin{equation}\label{eq: alternative representation of Hk}
H_k (y^n) = H_k (\mb(y^n)) :=\frac{1}{n} \sum_{\b,\bb} \mathcal{H} \left(
\mb_{\cdot,\bb}(y^n) \right) \mathbf{1}^T \mb_{\cdot,\bb}(y^n),
\end{equation}
where $\mathbf{1}$ and  $\mb_{\cdot,\bb}(y^n)$ denote the
all-ones column vector of length $|\hat{\Xc}|$, and the column
in $\mb(y^n)$ corresponding to $\bb$ respectively. For a vector
$\mathbf{v} = (v_1, \ldots , v_\ell)^T$ with non-negative
components, we let $\mathcal{H}(\mathbf{v})$ denote the entropy
of the random variable whose probability mass function (pmf) is
proportional to $\mathbf{v}$. Formally,
\begin{equation}\label{eq: single letter ent functional defined}
\mathcal{H} (\mathbf{v}) = \left\{ \begin{array}{cc}
                            \sum\limits_{i=1}^\ell \frac{v_i}{\| \mathbf{v} \|_1}  \log \frac{\| \mathbf{v} \|_1}{v_i} &  \mbox{ if }  \mathbf{v} \neq (0, \ldots , 0)^T \\
                            0 & \mbox{ if } \mathbf{v}  = (0, \ldots , 0)^T.
                          \end{array}
\right.
\end{equation}

\section{Linearized cost function} \label{sec: linearized cost}

Consider the following scheme for lossy source coding at fixed
slope $\a > 0$. For each source sequence $x^n$ let the
reconstruction block $\hat{x}^n$ be
\begin{equation}\label{eq: reconstruction via exhaustion for fixed slope s}
    \hat{x}^n = \mbox{arg} \min_{y^n\in\hat{\Xc}^n} \left[  H_k (y^n) +\a d_n (x^n,
    y^n)  \right].
\end{equation}
The encoder, after computing $\hat{x}^n$, losslessly conveys it
to the decoder using {\sf\footnotesize  LZ} compression. Let
$k$ grow slowly enough with $n$ so that
\begin{equation}\label{eq: how slowly need k grow with n}
  \limsup_{n \rightarrow \infty } \max_{y^n}  \left[ \frac{1}{n} \ell_{{\sf\footnotesize  LZ}} (y^n) -H_k
  (y^n) \right] \leq 0,
\end{equation}
where $\ell_{{\sf\footnotesize  LZ}} (y^n)$ denotes the length
of the {\sf\footnotesize  LZ} representation of $y^n$. Note
that Ziv's inequality guarantees that if $k = k_n = o(\log n)$
then \eqref{eq: how slowly need k grow with n} holds.
\begin{theorem} \cite{R_D MCMC} \label{th: the exhaustive search scheme}
Let $\mathbf{X}$ be a stationary and ergodic source, let
$R(\mathbf{X}, D)$ denote its rate distortion function,  and
let $\hat{X}^n$ denote the reconstruction using the above
scheme for coding $X^n$. Then
\begin{equation}\label{eq: achieving optimal point on the rd curve}
\mathds{E} \left[ \frac{1}{n} \ell_{{\sf\footnotesize  LZ}}
(\hat{X}^n) +\a  d_n (X^n,
    \hat{X}^n ) \right] \stackrel{n \rightarrow \infty}{\longrightarrow}  \min_{D \geq 0} \left[ R(\mathbf{X}, D)
    +\a D \right].
\end{equation}
\end{theorem}

In other words, conveying the reconstruction sequence to the
decoder via universal lossless compression (selection of
{\sf\footnotesize{LZ}}  algorithm here is for concreteness, but
other universal lossless methods  can  be used as well)
achieves optimum fixed-slope rate-distortion performance
universally.

As proposed in \cite{R_D MCMC}, the exhaustive search required
by this algorithm can be tackled through simulated annealing
Gibbs sampling. Here assuming the source is a discrete Markov
source, we propose another method for finding a sequence
achieving the minimum in \eqref{eq: reconstruction via
exhaustion for fixed slope s}. The advantage of the new method
is that its computational complexity is linear in $n$ for fixed
$k$.

Before describing the new scheme, consider the problems (P1)
and (P2) described below.

\begin{align}
(\textmd{P1}):\quad\min\limits_{y^n}\;\;\left[H_k(\mb(y^n))+\a
d_n(x^n,y^n)\right],
\end{align}
and
\begin{align}
(\textmd{P2}):\quad\min\limits_{y^n}\;\;\left[\sum\limits_{\b}\sum\limits_{\bb}
\l_{\b,\bb}m_{\b,\bb}(y^n)+\a  d_n(x^n,y^n)\right].
\end{align}

Comparing (P1) with \eqref{eq: reconstruction via exhaustion
for fixed slope s} reveals that it is the optimization required
by the exhaustive search coding scheme described before. The
question is whether it is possible to choose a set of
coefficients $\{\l_{\b,\bb}\}_{\b,\bb}$,
 $\b\in\hat{\Xc}$ and $\bb\in\hat{\Xc}^k$, such that (P1) and (P2) have
the same set of minimizers or at least, the set of minimizers
of (P2) is a subset of minimizers of (P1). If the answer to
this question is affirmative, then instead of solving (P1) one
can solve (P2), which, as we describe in Section \ref{sec:
Viterbi coding}, can be done simply via the Viterbi algorithm.

Let $\Sc_1$ and $\Sc_2$ denote the set of minimizers of (P1)
and (P2). Consider some $z^n\in \Sc_1$, and let
$\mb^*_n=\mb(z^n)$. Since $H(\mb)$ is concave in
$\mb$,\footnote{As proved in Appendix B.} for any empirical
count matrix $\mb$, we have
\begin{align}
H(\mb) &\leq H(\mb^*_n) + \sum\limits_{\b,\bb}
\frac{\partial}{\partial
m_{\b,\bb}}H(\mb)|_{\mb^*_n}(m_{\b,\bb}-m^*_{\b,\bb})\\
&\triangleq \hat{H}(\mb).
\end{align}
Now assume that in (P2), the coefficients are chosen as follows
\begin{equation}
\l_{\b,\bb}=\frac{\partial}{\partial
m_{\b,\bb}}H(\mb)\left|_{\mb^*_n}\right. .\label{eq: def of lambda}
\end{equation}

\begin{lemma}
(P1) and (P2) have the same minimum value, if the coefficients
are chosen according to \eqref{eq: def of lambda}. Moreover, if
all the sequences in $\Sc_1$ have the same type, then
$\Sc_1=\Sc_2$.
\end{lemma}
\begin{proof}
For any $y^n\in\hat{\Xc}^n$,
\begin{align}
H(\mb(y^n))+\a d_n(x^n,y^n)\leq \hat{H}(\mb(y^n))+\a d_n(x^n,y^n).
\end{align}
Therefore,
\begin{align}
\min\limits_{y^n}[H(\mb(y^n))+\a d_n(x^n,y^n)]&\leq
\min\limits_{y^n}[\hat{H}(\mb(y^n))+\a d_n(x^n,y^n)]\\
&\leq \hat{H}(\mb(z^n))+\a d_n(x^n,z^n)\\
&=\min\limits_{y^n}[H(\mb(y^n))+\a d_n(x^n,y^n)].
\end{align}
This shows that (P1) and (P2) have the same minimum values. For
any sequence $y^n$ with $\mb(y^n)\neq \mb^*_n$, by strict
concavity of $H(\mb)$,
\begin{align}
\hat{H}(\mb(y^n))+\a d_n(x^n,y^n)&>H(\mb(y^n))+\a d_n(x^n,y^n),\\
&\geq\min_{y^n} [H(\mb(y^n))+\a d_n(x^n,y^n)].
\end{align}
As a result all the sequences in $\Sc_2$ should have the
empirical count matrix equal to $\mb^*_n$. Since for these
sequences $H(\mb)=\hat{H}(\mb)$, we also conclude that
$\Sc_2\subset \Sc_1$. If there is a unique minimizing type
$\mb^*_n$, then $\Sc_1=\Sc_2$.
\end{proof}

This shows that if we knew the optimal type $\mb^*_n$, then we
could compute the optimal coefficients via \eqref{eq: def of
lambda}, and solve (P2) instead of (P1). The problem is that
$\mb^*_n$ is not known to the encoder (since knowledge of
$\mb_n^*$ requires solving (P1) which is the problem we are
trying to avoid). In the next section, we describe a method for
approximating $\mb^*_n$, and hence the coefficients
$\{\l_{\b,\bb}\}$.

\section{How to choose the coefficients?}\label{sec: how to choose}

For a given stationary ergodic source $\mathbf{X}$, and for any
given count matrix $\mb$ define $D(\mb)$ to be  the minimum
average expected distortion among all processes $\mathbf{Y}$
that are jointly stationary ergodic with $\mathbf{X}$ and their
$(k+1)^{\rm{th}}$ order stationary distribution is according to
$\mb$.\footnote{As discussed in Appendix A, the set of such
processes is non-empty for any legitimate $\mb$.} $D(\mb)$ can
equivalently be defined as
\begin{align}
D(\mb) = \lim\limits_{k_1\to\infty}
\min\limits_{p(x^{k_1},y^{k_1})\in\mathcal{M}^{(k_1)}}
\E_{p}d(x^{k_1},y^{k_1}),\label{eq: lim distortion}
\end{align}
where $\mathcal{M}^{(k_1)}$ is the set of all jointly
stationary distributions $p(x^{k_1},y^{k_1})$ of
$(X^{k_1},Y^{k_1})$ with marginal distributions with respect to
$x$ coinciding with the $k_1^{\rm{th}}$ order distribution of
$\mathbf{X}$ process, and with marginal distributions with
respect to $y$ coinciding with $\mb$, i.e., having the
$(k+1)^{\rm{th}}$ order marginal distribution described by
$\mb$.

\begin{lemma}\label{lemma: Markov source D(m)} If the source is $\ell^{\rm th}$ order Markov,
then
\begin{align}
D(\mb) = \min\limits_{p(x^{k_1},y^{k_1})\in\mathcal{M}^{(k_1)}}
\E_{p}d(x^{k_1},y^{k_1}),
\end{align}
where $k_1 = \max(\ell,k+1)$.
\end{lemma}

\begin{proof}[outline]
Using the technique described in Appendix A, for any legitimate
given joint distribution $p(x^{k1},y^{k1})$ with the marginal
distribution with respect to $x$ coinciding with the source
distribution and with marginal distribution with respect to $y$
coinciding with some given distribution $\mb$, it is possible
to construct a process which is jointly stationary and ergodic
with our source process and also has the $(k+1)^{\rm th}$ order
joint distribution as $p(x^{k1},y^{k1})$. Using this gives us
an achievable distortion, i.e., an upper bound on $D(\mb)$. On
the other hand, the limit given in \eqref{eq: lim distortion}
is approaching $D(\mb)$ from below. Combining the upper and
lower bounds yields the desired equality.
\end{proof}

Since by assumption the encoder does not know $\ell$, therefore
it can not compute $\max(\ell,k+1)$. But letting $k_1=k+1$,
where $k=o(\log n)$, for any fixed order $\ell$, $k_1$ will
eventually for $n$ large enough, exceed $\ell$, and hence be
equal to $\max(\ell,k+1)$. Having this observation in mind,
consider the following optimization problem,
\begin{align}
\min&\quad H(\mb)+\a D(\mb)\nonumber\\
\textmd{s.t.}&\hspace{4mm} \mb\in\mathcal{M}^{(k_1)}.\label{eq:
original optimization}
\end{align}
By Lemma \ref{lemma: Markov source D(m)}, an equivalent
representation of \eqref{eq: original optimization} is
\begin{align}
\min\quad H(\mb)&+\a \sum\limits_{\b,\b',\bb,\bb'} d_{k_1}(\b'\bb',\b\bb)p_x(\b'\bb')q_{y|x}(\b\bb|\b'\bb')\nonumber\\
\textmd{s.t.}\hspace{15mm} &m_{\b,\bb}\hspace{4mm}=\sum_{\b'\bb'}p_x(\b'\bb')q_{y|x}(\b \bb|\b'\bb'),\quad\forall\;\b,\bb,\nonumber\\
&0\leq  q_{y|x}(\b\bb|\b'\bb')\leq 1,\quad\forall\;\b,\b',\bb,\bb',\nonumber\\
&\sum_{\b,\bb}q_{y|x}(\b\bb|\b'\bb')=1,\quad \forall \; \b',\bb',\nonumber\\
&\sum\limits_{\b,\b'}p_x(\b'\bb')q_{y|x}(\b\bb|\b'\bb')=\sum\limits_{\b,\b'}p_x(\bb'\b')q_{y|x}(\bb\b|\bb'\b')\quad\forall\bb,\bb'.\label{eq: original Markov}
\end{align}
The last constraint in \eqref{eq: original Markov} is the
\emph{stationarity condition} defined in \eqref{eq: stat cond},
and ensures that the joint distribution defined by
$p_x(\b\bb)q_{y|x}(\b'\bb'|\b,\bb')$ over $(x^{k+1},y^{k+1})$
corresponds to $(k+1)^{\rm th}$ order marginal distribution  of
some jointly stationary processes $(\mathbf{X},\mathbf{Y})$.
Note that the variables in \eqref{eq: original Markov} are
conditional distributions $q_{y|x}(y^{k_1}|x^{k_1})$, but we
are only interested in the $\mb$ that they induce.

\begin{lemma}
If for each $n$, (P1) has a unique minimizing type $\mb^*_n$,
then
\begin{equation}
\|\mb^*_n-\hat{\mb}^*_n\|_{\rm TV}\to 0, \quad\quad{\rm a.s.},
\end{equation}
where $\hat{\mb}^*_n$ is the solution of \eqref{eq: original
Markov}.
\end{lemma}

\textbf{Remark:} In \eqref{eq: original Markov}, the only
dependence on $n$ is through $k_1$.

Therefore, if the encoder knew the distribution of the source,
it could solve \eqref{eq: original Markov}, find a good
approximation of $\mb^*_n$, and then use \eqref{eq: def of
lambda} to compute the coefficients required by (P2). The
problem is that the encoder does not have this information, and
only knows that the source is Markov (but does not know its
order). To overcome its lack of information, a reasonable step
is to use empirical distribution of the source instead of the
true unknown distribution in \eqref{eq: original Markov}. For
$a^{k_1}\in\Xc^{k_1}$, define the $k_1^{\rm th}$ order
empirical distribution of the source as
\begin{align} \label{eq: emp dist}
\hat{p}_x^{(k_1)}(a^{k_1})\triangleq\frac{|\{i:(x_{i-k_1},\ldots,x_{i-1})=a^{k_1}\}|}{n}.
\end{align}
The following lemma shows that for $k_1=o(\log n)$,
$\hat{p}^{(k_1)}$ converges to the actual $k_1^{\textmd{th}}$
order distribution of the source, and therefore can be
considered as a good approximation for it.
\begin{lemma}\label{lemma: convergence of emp counts}
For $k_1=o(\log n)$, and any stationary ergodic Markov source,
\begin{align}
\|\hat{p}^{(k_1)}-p^{(k_1)}\|_{\rm TV}\to 0 \quad\textmd{a.s.},
\end{align}
where $p^{(k_1)}$ is the true $k_1^{\rm th}$ order distribution
of the Markov source.
\end{lemma}

Assume $x^n$ is generated by a discrete Markov source, and let
$\hat{p}_x^{(k_1)}$ be its empirical distribution defined in
\eqref{eq: emp dist}. Consider the following optimization
problem
\begin{align}
\min\quad H(\mb)&+\a \sum\limits_{\b,\b',\bb,\bb'} d_{k_1}(\b'\bb',\b\bb)\hat{p}^{(k_1)}_x(\b'\bb')q_{y|x}(\b\bb|\b'\bb')\nonumber\\
\textmd{s.t.}\hspace{15mm} &m_{\b,\bb}\hspace{4mm}=\sum_{\b'\bb'}\hat{p}^{(k_1)}_x(\b'\bb')q_{y|x}(\b \bb|\b'\bb'),\quad\forall\;\b,\bb,\nonumber\\
&0\leq  q_{y|x}(\b\bb|\b'\bb')\leq 1,\quad\forall\;\b,\b',\bb,\bb'\nonumber\\
&\sum_{\b,\bb}q_{y|x}(\b\bb|\b'\bb')=1,\quad \forall \; \b',\bb',\nonumber\\
&\sum\limits_{\b,\b'}\hat{p}^{(k_1)}_x(\b'\bb')q_{y|x}(\b\bb|\b'\bb')=\sum\limits_{\b,\b'})\hat{p}^{(k_1)}_x(\bb'\b')q_{y|x}(\bb\b|\bb'\b'),\quad\forall\bb,\bb'.\label{eq: optimization}
\end{align}
and let $\tilde{\mb}_n^*$ denote the output of the above
optimization problem.

\begin{lemma}
For $k_1=k_1(n)=o(\log n)$,
\begin{align*}
\|\tilde{\mb}_n^*-\hat{\mb}_n^*\|_\textmd{TV}\to 0, \quad{a.s.}
\end{align*}
\end{lemma}
\begin{proof}[outline]
The input parameters of the optimization problem (\ref{eq:
optimization}) are
$\{\hat{p}^{(k_1)}(a^{k_1})\}_{a^{k_1}\in\Xc^{k_1}}$, therefore
$\hat{\mb}_n^*=\hat{\mb}_n^*(\{\hat{p}^{(k_1)}(a^{k_1})\}_{a^{k_1}\in\Xc^{k_1}})$.
On the other hand, both the cost function and the constraints
of \eqref{eq: optimization} are continuous both in input
parameters and optimization variables. This means that
$\hat{\mb}_n^*$ in turn is a continuous function of
$\{\hat{p}(x^{k_1})\}_{x^{k_1}\in\Xc^{k_1}}$.
\end{proof}

Let $\{\l_{\b,\bb}(n)\}_{\b,\bb}$ denote the optimal values of
the coefficients defined at $\mb_n^*$ (as given in \eqref{eq:
def of lambda}), and let $\{\hat{\l}_{\b,\bb}(n)\}_{\b,\bb}$ be
coefficients computed at $\tilde{\mb}_n^*$, then
\begin{lemma}
\begin{align}
\max\limits_{\b,\bb}|\l_{\b,\bb}(n)-\hat{\l}_{\b,\bb}(n)|\to 0
\textmd{ as
 } n\to\infty.
\end{align}
\end{lemma}

These results suggest that for computing the coefficients we
can solve the optimization problem given in \eqref{eq:
optimization} (whose complexity can be controlled with the rate
of increase of $k_1$), and then substitute the result in
\eqref{eq: def of lambda} to obtain the approximate
coefficients. After that (P2) defined by these coefficients can
be solved using the Viterbi algorithm in a way that will be
detailed in the next section. The succession of lemmas detailed
in the previous sections then allow us to prove the following
theorem.
\begin{theorem}\label{thm:2}
Let $\mathbf{X}$ let a stationary and ergodic Markov source,
and $R(\mathbf{X}, D)$ denote its rate distortion function. Let
$\hat{X}^n$ be the reconstruction sequence obtained using the
above scheme for coding $X^n$ choosing $k_1=k+1$, where
$k=o(\log n)$. Then
\begin{equation}\label{eq: achieving optimal point on the rd curve}
\mathds{E} \left[ \frac{1}{n} H_{k}(\bold{m}(\hat{X}^n)) +\a  d_n (X^n,
    \hat{X}^n ) \right] \stackrel{n \rightarrow \infty}{\longrightarrow}  \min_{D \geq 0} \left[ R(\mathbf{X}, D)
    +\a D \right].
\end{equation}
\end{theorem}

{\bf Remark:} Theorem  \ref{thm:2} implies the fixed-slope
universality of the scheme which does the lossless compression
of the reconstruction by first describing its count matrix
(costing a number of bits which is negligible for large $n$)
and then doing the conditional entropy coding.
%

\section{Viterbi coder}\label{sec: Viterbi coding}

As proved in Section \ref{sec: linearized cost}, instead of
solving (P1), one can solve (P2) for proper choices of
coefficients $\{\l_{\bb,\b}\}$. Note that
\begin{align}
&\sum\limits_{\b,\bb}\left[
\l_{\b,\bb}m_{\b,\bb}(y^n)+\a
d_n(x^n,y^n)\right]=\frac{1}{n}\sum\limits_{i=1}^n\left[
\l_{y_i,y_{i-k}^{i-1}}+\a d(x_i,y_i)\right].\label{eq:3}
\end{align}
This alternative representation of the cost function suggests
that instead of using simulated annealing, we can find the
sequence that minimizes the cost function by the Viterbi
algorithm. For $i=k+1,\ldots,n$, let $s_i=y_{i-k}^i$ be the
state at time $i$, $\cal{S}$ be the set of all $2^{k+1}$
possible states, and for $s=b^{k+1}$ define
\[
w(s,i):=\l_{b_{k+1},b^k}+\a d(x_i,b_{k+1}).\] From our
definition of the states $s_{i}=g(s_{i-1},y_i)$, where
$g:\Sc\times\hat{\Xc}\to\Sc$. This representation leads to a
Trellis diagram corresponding to the evolution of the states
$\{s_i\}_{i=k+1}^m$ in which each state has $|\hat{\Xc}|$
states leading to it and $|\hat{\Xc}|$ states branching from
it. Assume that weight $w(s_i,i)$  is assigned to the edge
connecting states $s_{i-1}$ and $s_i$, i.e., the cost of each
edge only depends on the tail state.

It is clear that in our representation,  there is a 1-to-1
correspondence between sequences $y^n\in\hat{\Xc}^n$ and
sequences of states $\{s_i\}_{i=k+1}^m$, and minimizing
\eqref{eq:3} is equivalent to finding the path of minimum
weight in the corresponding Trellis diagram, i.e., the path
$\{s_i\}_{i=k+1}^n$ that minimizes $\sum_{i=k+1}^n w(s_i,i)$.
Solving this minimization can readily be done by Viterbi
algorithm which can be described as follows. For each state
$s$, let $\mathcal{L}(s)$ be the two states leading to it, and
for any $i>1$,
\begin{align}
C(s,i):=\min\limits_{s'\in\mathcal{L}(s)}[w(s,i)+C(s',i-1)].
\end{align}
For $i=1$ and $s=b^{k+1}$, let $C(s,1):=\l_{b_{k+1},b^k}+\a
d_{k+1}(x^{k+1},b^{k+1})$. Using this procedure, each state $s$
at each time $j$ has a path of length $j-k-1$ which is the
minimum path among all the possible paths between $i=k+1$ and
$i=j$ such that $s_j=s$. After computing
$\{C(s,i)\}_{\substack{s\in\Sc\\i\in\{k+1,\ldots,n\}}}$, at
time $i=n$, let
\begin{align}
s^*=\argmin_{s\in\Sc}C(s,n).
\end{align}

It is not hard to see  that the path leading to $s^*$ is the
path of minimum weight among all possible paths.

Note that the computational complexity of this procedure is
linear in $n$ but exponential in $k$ because the number of
states increases exponentially with $k$.
\begin{figure}[t]
\begin{center}
\includegraphics[width=11cm]{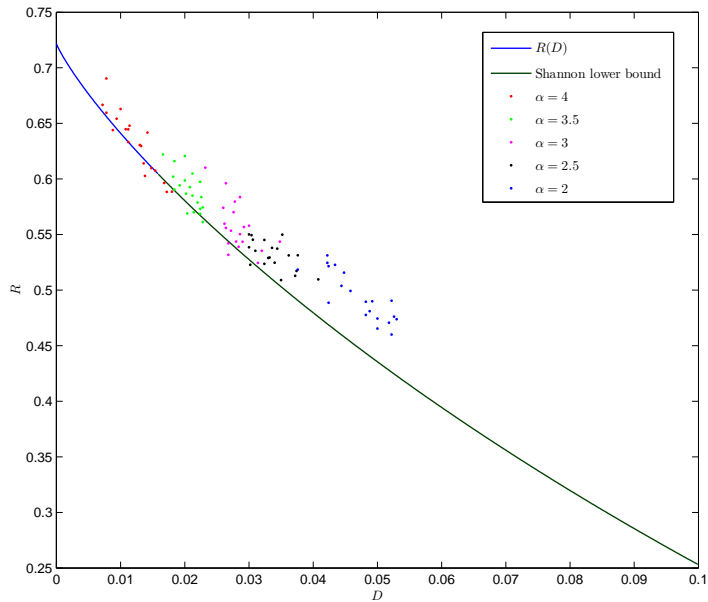}
\vspace{-0.8cm}\caption{ $(d_n(x^n,\hat{x}^n),H_k(\hat{x}^n))$ of
output points of Viterbi encoder when the coefficients are computed
at $\mb[x^n]$. For each value of $\a$, the algorithm is run $L=20$
times. Here $n=5000$, $k=7$, and the source is binary Markov with
$q=0.2$}\label{fig:1}
\end{center}
\end{figure}

\begin{figure}
\begin{center}
\includegraphics[width=11cm]{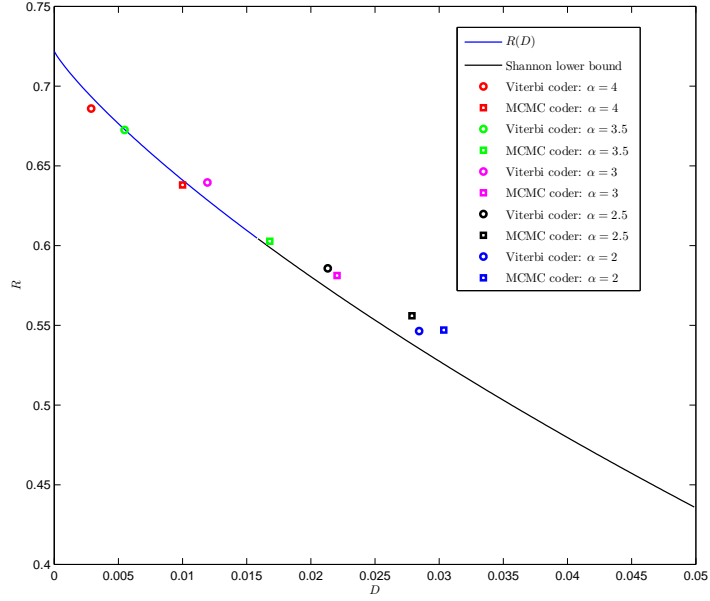}
\vspace{-0.8cm}\caption{Comparing the performances of Viterbi
encoder and MCMC encoder proposed in \cite{R_D MCMC}}\label{fig:3}
\end{center}
\end{figure}

\section{Simulation results}\label{sec: simulation results}
In this section, some preliminary simulation results of the
application of Viterbi encoder described in the previous
section is presented. In our simulations, instead of computing
the coefficients $\{\l_{\b,\bb}\}$ from \eqref{eq: def of
lambda} at the optimal point $\mb_n^{*}$, we compute them at
the count matrix of the input sequence $x^n$, $\mb(x^n)$.
Fig.~\ref{fig:1} demonstrates $(d_n(x^n,y^n),H_k(\mb(y^n)))$ of
output points of the described algorithm. The block length is
$n=5000$, $k=7$ and the source is $1^{\textmd{st}}$ order
binary symmetric Markov with transition probability $q=0.2$.
For each value of $\alpha$ the algorithm is applied to $L=20$
different randomly generated sequences. The reason of getting
some points below the rate-distortion curve is that the actual
number of bits required for describing $\hat{x}^n$ losslessly
to the decoder is larger than $nH_k(\hat{x}^n)$, but converges
to it as $n$ grows. For example, for the simple scheme of
separately describing the subsequences corresponding to
different preceding contexts, this surplus is of order $2^k\log
n/n$. The effect of this excess rate is not reflected in the
figure, which explains why some points appear below the
rate-distortion curve.

It can be observed that for larger values of $\a$ the output
points are closer to the curve. The reason is that large values
of $\a$ correspond to small values of distortion, and if the
distortion is small then $\mb(x^n)$ is a good approximation of
$\mb(y^n)$.


Finally, Fig.~\ref{fig:3} compares the performance of the new
Viterbi encoder and the MCMC encoder described in \cite{R_D
MCMC}. Here the source is again binary symmetric Markov with
$q=0.2$, and the other parameters are: $k=7$, $n=5,000$,
$\beta_t=n\log t$, $r=10n$, where $\beta_t$ determines the
cooling schedule of the MCMC coder and $r$ is its number of
iterations. Each point is the figure corresponds to the average
performance of $L=10$ random realizations of the source. It can
be observed that even for this simplistic choice of the
coefficients the performance of the algorithms are comparable,
while the Viterbi encoder for example in this example runs at
least $40$ times faster.

\section{Conclusions and Current Directions}\label{sec: conclusion}

In this paper, a new method for universal fixed-slope lossy
compression of discrete Markov sources was proposed. The new
method achieves the rate-distortion curve for any discrete
Markov source. Extending the algorithm to work on any
stationary ergodic source is under current investigation. We
believe that in fact the same algorithm works for the general
class of stationary ergodic sources, and only the proof should
be extended to work in this case as well. Another direction for
future work is finding a simple method for approximating the
optimal coefficients that would alleviate the need for solving
the optimization problem \eqref{eq: optimization}.

\renewcommand{\theequation}{A-\arabic{equation}}
\setcounter{equation}{0}  

\section*{APPENDIX A: Stationarity  condition}  \label{app1} 

Assume that we are given a $|\hat{\Xc}|\times |\hat{\Xc}|^k$ matrix $\mb$ with all
elements positive and summing up to one. The question is under what condition(s) this matrix can be $(k+1)^{\rm th}$ order stationary distribution of a stationary process. For the ease of notations, instead of matrix $\mb$ consider $p(x^{k+1})$ as a distribution defined on $\hat{\Xc}^{k+1}$. We show that a necessary and sufficient condition is the so-called \emph{stationarity condition} which is
\begin{align}
\sum\limits_{\b\in\hat{\Xc}}p(\b x^k)=\sum\limits_{\b\in\hat{\Xc}}p(x^k\b). \label{eq: stat cond}
\end{align}
\begin{itemize}
\item[-] Necessity: The necessity of \eqref{eq: stat cond}
    is just a direct result of the definition of
    stationarity of a process. If $p(x^{k+1})$ is to
    represent the $(k+1)^{\rm th}$ order marginal
    distribution of a stationary process, then it should be
    consistent with the $k^{\rm th}$ order marginal
    distribution as satisfy \eqref{eq: stat cond}.
\item[-] Sufficiency: In order to prove the sufficiency, we
    assume that \eqref{eq: stat cond} holds, and build a
    stationary process with  $(k+1)^{\rm th}$ order
    marginal distribution of $p(x^{k+1})$. Consider a
    $k^{\rm th}$ order Markov chain with transition
    probabilities of
\begin{align}
q(x_{k+1}|x^k)=\frac{p(x^{k+1})}{p(x^k)}.
\end{align}
Note that $p(x^k)$ is well-defined by \eqref{eq: stat cond}. Moreover, again from \eqref{eq: stat cond}, $p(x^{k+1})$ is the stationary distribution of the defined Markov chain, because
\begin{align}
\sum\limits_{x_1}q(x_{k+1}|x^k)p(x^k)=\sum\limits_{x_1}p(x^{k+1})=p(x_2^{k+1}).
\end{align}
Therefore  we have found a stationary process that has the desired marginal distribution.


\end{itemize}

Finally we show that if $\mb$ is the count matrix of a sequence $y^n$, then there exist a stationary process with the marginal distribution coinciding with $\mb$. From what we just proved, we only need to show that \eqref{eq: stat cond} holds, i.e.,
\begin{align}
\sum\limits_{\b}m_{\b,\bb}=\sum\limits_{\b}m_{b_k,[\b,b_1\ldots,b_{k-1}]}. \label{eq: m stat}
\end{align}
But this is true because both sides of \eqref{eq: m stat} are equal to $|\{i:y_{i+1}^{i+k}=\bb\}|/n$.

\renewcommand{\theequation}{B-\arabic{equation}}
\setcounter{equation}{0}  

\section*{APPENDIX B: Concavity of $H(\mb)$ }  \label{app2} 
For simplicity assume that $\Xc=\hat{\Xc}=\{0,1\}$. By
definition
\begin{align}
H(\mb)=\sum\limits_{\bb\in\{0,1\}^k}(m_{0,\bb}+m_{1,\bb})h(\frac{m_{0,\bb}}{m_{0,\bb}+m_{1,\bb}}),
\end{align}
where $h(\a)=\a\log{\a}+\bar{\a}\log{\bar{\a}}$ and
$\bar{\a}=1-\a$. We need to show that for any $\theta\in[0,1]$,
and empirical count matrices $\mb^{(1)}$ and $\mb^{(2)}$,
\begin{align}
\theta H(\mb^{(1)})+\bar{\theta}H(\mb^{(2)})\leq
H(\theta\mb^{(1)}+\bar{\theta}\mb^{(2)}).
\end{align}
From the concavity of $h$, it follows that
\begin{align}
&\theta(m^{(1)}_{0,\bb}+m^{(1)}_{1,\bb})h(\frac{
m^{(1)}_{0,\bb}}{m^{(1)}_{0,\bb}+m^{(1)}_{1,\bb}})+\bar{\theta}(m^{(2)}_{0,\bb}+m^{(2)}_{2,\bb})h(\frac{
m^{(2)}_{0,\bb}}{m^{(2)}_{0,\bb}+m^{(2)}_{2,\bb}})\nonumber\\
&=(\theta(m^{(1)}_{0,\bb}+m^{(1)}_{1,\bb})+\bar{\theta}
(m^{(2)}_{0,\bb}+m^{(2)}_{1,\bb}))\sum\limits_{i\in\{1,2\}}\frac{\theta_i(m^{(i)}_{0,\bb}+m^{(i)}_{1,\bb})}{(\theta
(m^{(1)}_{0,\bb}+m^{(1)}_{1,\bb})+\bar{\theta}
(m^{(2)}_{0,\bb}+m^{(2)}_{1,\bb}))}h(\frac{
m^{(i)}_{0,\bb}}{m^{(i)}_{0,\bb}+m^{(i)}_{1,\bb}})\nonumber\\
&\leq (\theta(m^{(1)}_{0,\bb}+m^{(1)}_{1,\bb})+\bar{\theta}
(m^{(2)}_{0,\bb}+m^{(2)}_{1,\bb})) h(\frac{ \theta m^{(1)}_{0,\bb}+
\bar{\theta}m^{(2)}_{0,\bb}}{\theta
(m^{(1)}_{0,\bb}+m^{(1)}_{1,\bb})+\bar{\theta}
(m^{(2)}_{0,\bb}+m^{(2)}_{1,\bb})}),\label{eq: concavity}
\end{align}
where $\theta_1=1-\theta_2=\theta$. Now summing up both sides
of \eqref{eq: concavity} over all $\bb\in\hat{\Xc}^{k}$, yields
the desired result.

\end{document}